\documentclass[11pt]{article}
\usepackage{amsmath,amssymb,epsfig,bm}

\date{\empty}

\begin{document}

\title{\bf Causality, initial conditions, and inflationary magnetogenesis}

\author{Christos G. Tsagas\\ {\small Section of Astrophysics, Astronomy and Mechanics, Department of Physics}\\ {\small Aristotle University of Thessaloniki, Thessaloniki 54124, Greece}}

\maketitle

\begin{abstract}
The post-inflationary evolution of inflation-produced magnetic fields, conventional or not, can change dramatically when two fundamental issues are accounted for. The first is causality, which demands that local physical processes can never affect superhorizon perturbations. The second is the nature of the transition from inflation to reheating and then to the radiation era, which determine the initial conditions at the start of these epochs. Causality implies that inflationary magnetic fields dot not freeze into the matter until they have re-entered the causal horizon. The nature of the cosmological transitions and the associated initial conditions, on the other hand, determine the large-scale magnetic evolution after inflation. Put together, the two can slow down the adiabatic magnetic decay on superhorizon scales throughout the universe's post-inflationary evolution and thus lead to considerably stronger residual magnetic fields. This is ``good news'' for both the conventional and the non-conventional scenarios of cosmic magnetogenesis. Mechanisms operating outside standard electromagnetism, in particular, do not need to enhance their fields too much during inflation, in order to produce seeds that can feed the galactic dynamo today. In fact, even conventionally produced inflationary magnetic fields might be able to sustain the dynamo.
\end{abstract}

\section{Introduction}\label{sI}
Despite the efforts, the quest for the origin of the large-scale magnetic ($B$) fields seen in the universe today still goes on~\cite{K}. Two are the main ``schools of thought''. The first suggests a late-time generation for these fields, triggered by physical processes that operate after recombination, while the second advocates a primordial origin for cosmic magnetism. In both cases, the aim is to generate the seeds that will be later amplified by the galactic dynamo to produce the large-scale $B$-fields observed in galaxies~\cite{Kr}. To operate successfully, however, the dynamo requires seeds with certain specifications. One is the coherence scale, which should not be less than 10~Kpc (comoving, namely before the collapse of the protogalaxy). The other requirement is the seed's strength, which depends on the efficiency of the dynamo amplification and typically varies between $10^{-22}$ and $10^{-12}$ Gauss. Note that these magnitudes are measured at the time of completed galaxy formation, that is after the collapse of the protogalaxy. The presence of $B$-fields in high-redshift protogalaxies, where the dynamo had less time to operate, with strengths similar to that of their galactic counterparts (of $\mu$G-order), could be interpreted as a sign in favour of primordial magnetism~\cite{GR}. This idea has received additional boost from recent reports suggesting the existence of magnetic fields close to $10^{-15}$~G in empty intergalactic space, where presumably no dynamo mechanism can operate~\cite{NV}. Nevertheless, producing $B$-fields in the early universe that will successfully seed the galactic dynamo today has proved anything but straightforward.

Assuming that the magnetic seeds are generated after inflation, the main problem is their coherence length, which is generally much smaller than the required 10~Kpc. This is due to causality, which confines the scale of any newly produced $B$-field inside the horizon at the time. The latter is typically too small. Theoretically, one could address the size-issue by appealing to ``inverse cascade'', a mechanism that can transfer magnetic energy from smaller to larger scales and thus increase the effective length of the field~\cite{BEO}. The jury is still out, however, as it appears that inverse cascade requires rather large amounts of magnetic helicity in order to operate efficiently. There is no size-problem whatsoever for the inflationary magnetic fields, since inflation naturally achieves very large correlation lengths. Here, the main obstacle is the residual strength of the field, which is believed to be too weak (less than $10^{-50}$~G at present) to have any astrophysical significance. This extreme weakness has been largely attributed to the so-called adiabatic magnetic decay. The belief, in other words, is that magnetic fields deplete as $B\propto a^{-2}$ (where $a$ is the cosmological scale factor) throughput their evolution and on all scales.

\section{Magnetic fields with superhorizon
correlations}\label{sMFSCs}
The adiabatic magnetic decay is attributed to the rapidly increasing electrical conductivity of the post-inflationary universe, which is thought to guarantee that $B$-fields remain frozen into the cosmic medium at all times and on all scales. However, the magnetic freeze-in cannot take place without the electric currents. These are generated after inflation and therefore, unlike the inflation-produced $B$-fields, their size is always confined within the causal horizon. Put another way, there can be not electric currents with super-Hubble correlations. Moreover, the freeze-in process itself is causal and therefore it cannot affect magnetic fields larger than the horizon at the time. Arguing for the opposite is claiming that local causal physics can affect superhorizon-sized perturbations, which openly violates causality (e.g.~see~\cite{RW} for related quotes). All these mean that inflationary magnetic fields do not necessarily freeze-in until they have crossed back inside the horizon and have come again into full causal contact.\footnote{Once back inside the causal horizon, the electric currents quickly freeze the magnetic fields into the highly conductive matter of the post-inflationary universe. Then after, the $B$-fields decay adiabatically to the present.} Before horizon entry, these $B$-fields were causally disconnected and therefore immune to local physics, being affected by the universal expansion only. At the same time, like any other superhorizon perturbation produced during inflation, the aforementioned magnetic fields retain the memory of their de Sitter past. In practice, this means that, as long as they remain outside the Hubble radius of a spatially flat Friedmann-Robertson-Walker (FRW) universe, $B$-fields with superhorizon correlations obey the long-wavelength solution
\begin{equation}
\mathcal{B}\equiv a^2B= C_1+ C_2n\eta\,,  \label{cB}
\end{equation}
of the linear ``source-free'' wave-like formula $\mathcal{B}^{\prime\prime}-a^2{\rm D}^2\mathcal{B}=0$~\cite{GR}. Note that $\mathcal{B}=a^2B$ is the rescaled magnetic field, $n$ is its comoving wavenumber (with $n>0$), $\eta$ is the conformal time and primes indicate conformal-time derivatives. Also, since the $B$-fields in question are well outside the Hubble radius, they satisfy the constraint $n\eta\ll1$. The latter has led many authors to disregard the second mode on the right-hand side of solution (\ref{cB}) and thus conclude that magnetic fields decay adiabatically on super-Hubble scales as well. It is conceivable, however, that there are initial conditions allowing for $C_2\gg C_1$, in which case the aforementioned ``redundant'' mode can dominate. After all, this is an essentially ``growing'' mode and for this reason it should not be a priori ignored, at least not before the integration constants have been evaluated.

Before discussing the implications of solution (\ref{cB}), we should remind the reader that the direct link between the aforementioned source-free treatment of superhorizon-sized magnetic fields and causality was originally made in~\cite{BT}. Those studies, however, focussed on spatially open Friedmann-Robertson-Walker (FRW) universes. The first extended discussion of the causality issue and of its potentially pivotal implications for cosmic magnetogenesis in spatially flat FRW universes was given in~\cite{T1}. That work was then specialised to non-conventional scenarios of inflationary magnetic generation in~\cite{T2}. In both cases, the final strength of the aforementioned large-scale $B$-fields was found to be much larger than the typical values quoted in the standard literature. More recently, the same source-free approach to the study of superhorozon-sized magnetic fields was also adopted in~\cite{C1,C2}, this time from the viewpoint of high energy physics, with analogous results. Here, we will concentrate on the role and the consequences of the initial conditions for the post-inflationary evolution of large-scale primordial (conventional or not) magnetic fields.

Let us now go back to Eq.~(\ref{cB}) and to the integration constants seen there. These are fairly straightforward to calculate and once this is done solution (\ref{cB}) reads~\cite{T1}
\begin{equation}
B= [B_*-\eta_*(2a_*H_*B_*+B_*^{\prime})]\left({a_*\over a}\right)^2+ \eta_*(2a_*H_*B_*+B_*^{\prime})\left({a_*\over a}\right)^2\left({\eta\over\eta_*}\right)  \label{B1}
\end{equation}
where $H=a^{\prime}/a^2$ is the Hubble parameter and the $\star$-suffix marks the start of a post-inflationary cosmological epoch (e.g.~the reheating, the radiation or the dust era). Also, $a=a(\eta)$ with $a\geq a_*$ and $\eta\geq\eta_*$. Recalling that after inflation the conformal time is proportional to a positive power of the scale factor (i.e.~$a\propto\eta^k$ with $k>0$), it becomes immediately obvious that the second mode of (\ref{B1}) decays slower that its adiabatic counterpart. Whether this slowly decaying mode survives or not depends on the initial conditions, which determine its coefficient. All these make the post-inflationary evolution of superhorizon-sized magnetic fields a matter of initial conditions. At the start of reheating, the latter are decided by the magnetic evolution during inflation and by the nature of the transition from inflation to reheating. The initial conditions at the beginning of the subsequent epochs are determined in an analogous way as well. In what follows we will consider three complementary sets of initial conditions, keeping in mind that alternative scenarios may also be possible.

\section{Initial conditions and alternative 
scenarios}\label{sICASs}
In the literature, the transitions from inflation to reheating and then to the radiation era are generally treated as discontinuous, which in a sense reflects the overall uncertainty still clouding these events. More specifically, in both cases the equation of state of the universe is allowed to undergo an abrupt change. As a result, one needs to ``match'' the spacetime prior to the transition with the one after and the standard way of doing this is by appealing to Israel's junction conditions~\cite{I}. According to these, the matching of the aforementioned two spacetimes depends on whether the 3-dimensional transition hypersurface ($\Sigma$) has zero or finite ``width''. In the former case there is no ``thin shell'' on the matching hypersurface, while in the latter there is a thin ``surface layer'' with a finite stress-energy tensor (e.g.~see~\cite{P}). The main difference between the two alternatives is that, in the absence of thin shells, there can be no discontinuity in the extrinsic curvature on $\Sigma$. When dealing with an FRW background, this demands that the value of the Hubble parameter remains the same before and after the transition. More specifically, $[H_*]_-^+=H_*^+-H_*^-=0$, with the ``$-$'' and ``$+$'' superscripts denoting the moments just prior and immediately after the transition respectively.\footnote{Hereafter, the $\star$-suffix will always indicate the moment of the transition from one cosmological epoch to the next, while the zero suffix will correspond to the present (see Eqs.~(\ref{B5}) and (\ref{B8}) below).} In the presence of thin surface layers, on the other hand, there is a ``jump'' in the Hubble value measured on either side of the transit surface (i.e.~$[H_*]_-^+=H_*^+-H_*^-\neq0$). Note that in the former case $\Sigma$ is the hypersurface of constant density, but not necessarily of constant time (e.g.~see~\cite{DM}), while in the latter $\Sigma$ is typically the hypersurface of constant (absolute) conformal time (e.g.~see~\cite{CW}). Finally, there can be no discontinuity in the cosmological scale factor in either case, which ensures that $[a_*]_-^+=a_*^+-a_*^-=0$ always (e..g.~see~\cite{DM,CW}).

Assuming that the nature of the transition from inflation to reheating and then to the radiation era is ``fixed'' by Israel's junction conditions, the remaining degree of freedom is decided by the magnetic evolution during the de Sitter phase. Here, we will consider two alternatives. In the first, we will assume that the $B$-field decayed adiabatically (i.e.~that $B\propto a^{-2}$) throughout inflation, as it is the case for the conventional fields. The second alternative, on the other hand, will allow for the superadiabatic amplification of the magnetic field. To be precise, we will assume that $B\propto a^{-m}$ (with $0<m<2$) all along the de Sitter expansion, as it happens in typical non-conventional scenarios of inflationary magnetogenesis (e.g.~see~\cite{S}).\vspace{3mm}

\textit{Scenario A:} Consider conventional inflationary magnetic fields, produced within the framework of standard electromagnetism, which had been decaying adiabatically throughout the de Sitter regime. Suppose also that there is no thin shell on the transition hypersurface from inflation to reheating. Recalling that $H=a^{\prime}/a^2$, we deduce that $B_*^{\prime\;-}= -2a_*^-H_*^-B_*^-$ at the end of inflation. Then, given that $a_*^+=a_*^-$ always, that $H_*^+=H_*^-$ (in the absence of surface layers) and assuming that there is no discontinuity in the magnetic field on the matching surface (i.e.~setting $B_*^+=B_*^-$ and $B_*^{\prime\;+}=B_*^{\prime\;-}$ on either side of $\Sigma$), we obtain $B_*^{\prime\;+}= -2a_*^+H_*^+B_*^+$ at the start of reheating~\cite{T1}. On using these initial conditions, the coefficient of the second mode on the right-hand side of solution (\ref{B1}) vanishes, leaving the adiabatically decaying mode as the sole survivor. As a result, throughout the reheating phase we have
\begin{equation}
B= B_*^+\left({a_*^+\over a}\right)^2\,,  \label{B2}
\end{equation}
where $a\geq a_*^+$. It is straightforward to verify that the situation repeats itself at the transition to radiation epoch and later to that of dust, as long as there is no discontinuity in the value of the Hubble parameter at the time of the transit~\cite{T1}. Overall, one could argue that, in the absence of surface layers on the hypersurfaces connecting consecutive cosmological epochs, large-scale magnetic fields that happen to decay adiabatically prior to the transition, will keep doing so throughout the subsequent era as well. Thus, for all practical purposes, the initial conditions adopted in this scenario have reproduced the standard story of conventional inflationary magnetogenesis. The latter leads to astrophysically irrelevant magnetic fields with residual strengths around $10^{-53}$~G or less (on comoving scales of $\sim10$~Kpc or more).\vspace{3mm}

\textit{Scenario B:} Most mechanisms of non-conventional inflationary magnetogenesis operate outside standard electromagnetic theory, with the latter typically restored once the de Sitter phase is over. Breaking away from Maxwellian electromagnetism can lead to strong superadiabatic amplification and thus achieve magnetic fields that, by the end of inflation, are strong enough to seed the galactic dynamo today, despite their post-inflationary adiabatic decay. There are caveats, however, one of which is the so-called ``backreaction problem'', where the magnetic enhancement is so efficient that it starts interfering with the dynamics of the inflationary expansion (e.g.~see~\cite{DMR}). However, producing very strong magnetic fields by the end of de Sitter regime is not necessary, provided the subsequent magnetic decay is slower than the adiabatic. To demonstrate this, let us consider a $B$-field that had been decaying superadiabatically, namely as $B\propto a^{-m}$ with $0<m<2$, all along inflation. At the end of that period we have $B_*^{\prime\;-}= -ma_*^-H_*^-B_*^-$, which translates into $B_*^{\prime\;+}= -ma_*^+H_*^+B_*^+$ at the start of reheating (recall that $[a_*]_-^+=0=[H_*]_-^+=[B_*]_-^+=[B_*^{\prime}]_-^+$ similarly to Scenario~A above). Therefore, at the onset of the reheating epoch, solution (\ref{B1}) reads
\begin{equation}
B= [1-(2-m)\eta_*^+a_*^+H_*^+]B_*^+\left({a_*^+\over a}\right)^2+ (2-m)\eta_*^+a_*^+H_*^+B_*^+\left({a_*^+\over a}\right)^2\left({\eta\over\eta_*^+}\right)\,,  \label{B3}
\end{equation}
with $a\geq a_*^+$ and $\eta\geq\eta_*^+>0$. Moreover, recalling that $a\propto\eta^2$ and $H=2/a\eta$ throughout reheating, the above reduces to~\cite{T1,T2}
\begin{equation}
B= -(3-2m)B_*^+\left({a_*^+\over a}\right)^2+ 2(2-m)B_*^+\left({a_*^+\over a}\right)^{3/2}\,.  \label{B4}
\end{equation}
Given that $m\neq2$, the slowly decaying second mode on the right-hand side of (\ref{B4}) survives and the $B$-field is superadiabatically amplified all along the reheating era. In the same way, one can show that the superadiabatic amplification persists into the radiation and the dust epochs, with $B\propto a^{-1}$ and $B\propto a^{-3/2}$ respectively, as long as the $B$-field remains outside the Hubble radius~\cite{T1,T2}. As mentioned before, once back inside the horizon, the highly conductive electric currents will freeze the magnetic field into the matter and thus ``restore'' the adiabatic decay-law. In practice, all these mean that a $B$-field with current comoving size close to 10~Kpc (the minimum required by the galactic dynamo) and strength $B_{DS}$ at the end of the de Sitter expansion will have residual magnitude
\begin{equation}
B_0\simeq B_{DS}\,{T_0^2T_{RH}\over T_{HC}M^2}\,,  \label{B5}
\end{equation}
today~\cite{T2}. Note that $T_0$ is the current temperature of the universe, $T_{RH}$ is the reheat temperature, $T_{HC}$ is the temperature at the time the magnetic field crossed back inside the Hubble radius and $M$ is the energy scale of the adopted inflationary model (all measured in GeV). Wavelengths close to 10~Kpc today have re-entered the horizon prior to equipartition at $T_{HC}\simeq10^{-6}$~GeV (recall that $T_{EQ}\simeq10^{-9}$~GeV). Consequently, setting $T_0\simeq10^{-13}$~GeV, $T_{RH}\simeq10^{10}$~GeV and $M\simeq10^{17}$~GeV, we obtain $B_0\simeq10^{-44}B_{DS}$. This is the comoving value of the field, calculated before the collapse of the protogalaxy. When the latter is anisotropic (see~\cite{DBL} for related studies), the magnetic strength can increase by up to six orders of magnitude to $B_0\simeq10^{-38}B_{DS}$. Magnetic fields are capable of seeding the galactic dynamo when $B_0\gtrsim10^{-22}$~G, which in our case is achieved when $B_{DS}\gtrsim10^{16}$~G at the end of the de Sitter regime. Strengths of this magnitude are fairly easy to achieve by non-conventional mecamisms of inflationary magnetogenesis, without causing any backreaction or other known problems (e.g.~see~\cite{KSW}). Note that, if magnetic fields were to decay adiabatically on all scales after inflation, the minimum required strength at the end of the de Sitter phase would have exceeded $10^{40}$~G.\vspace{3mm}

\textit{Scenario C:} Let us now assume that the $B$-fields decay adiabatically during the de Sitter regime, but that there is a finite surface layer (a thin shell) on the transit hypersurface ($\Sigma$) to reheating. Then, following Israel's junction conditions, there is a jump in the value of the Hubble parameter at the time of the transition. This ensures that $H_*^+\neq H_*^-$, which implies that the coefficient of the second mode on the right-hand side of (\ref{B1}) is not necessarily zero. Consequently, the adiabatic magnetic decay after the transition (see scenario~A) is no longer guaranteed. To show this recall that during reheating the cosmological scale factor and the conformal time are related by $a\propto\eta^2$, with $\eta>0$, while the Hubble parameter is $H=2/a\eta$. Then, at the start of reheating, solution (\ref{B1}) recasts into
\begin{equation}
B= -(3B_*^++\eta_*^+B_*^{\prime\;+})\left({a_*^+\over a}\right)^2+ (4B_*^++\eta_*^+B_*^{\prime\;+})\left({a_*^+\over a}\right)^{3/2}\,,  \label{B6}
\end{equation}
where $a\geq a_*^+$. Assuming that the $B$-field decayed adiabatically during the de Sitter phase, we have $B_*^{\prime\;-}=-2a_*^-H_*^-B_*^-$ at the end of inflation proper (see also Scenario~A before). The aforementioned condition is also written as $\eta_*^-B_*^{\prime\;-}=2B_*^-$, given that $a\propto-1/\eta$, with $\eta<0$, and $H=-1/a\eta$ throughout the inflationary expansion. Therefore, setting $\eta_*^+=-\eta_*^-$ (recall that $\eta_*^-<0$ and $\eta_*^+>0$), $B_*^+=B_*^-$ and $B_*^{\prime\;+}=B_*^{\prime\;-}$, we have $\eta_*^+B_*^{\prime\;+}=-2B_*^+$ at the beginning of reheating. Substituting these initial conditions to the right-hand side of (\ref{B2}) we arrive at
\begin{equation}
B= -B_*^+\left({a_*^+\over a}\right)^2+ 2B_*^+\left({a_*^+\over a}\right)^{3/2}\,.  \label{B7}
\end{equation}
which shows that the $B$-field no longer decays adiabatically~\cite{T1}. Instead, the magnetic decay rate has slowed down to $B\propto a^{-3/2}$, which means that the field is superadiabatically amplified throughout the reheating era. Proceeding in an exactly analogous way, it is straightforward to demonstrate that the magnetic superadiabatic amplification continues into the radiation and the dust epochs, as long as the field remains larger than the Hubble radius. In particular, superhorizon-sized fields decay as $B\propto a^{-1}$ and $B\propto a^{-3/2}$, during the radiation and the dust eras respectively~\cite{T1,T2}. Scenario~C affects conventional inflation-produced magnetic fields and allows them to achieve residual magnitudes much stronger than those typically quoted in the literature. For example, a $B$-field with current scale close to 10~Kpc that crossed back inside the horizon in the late radiation epoch, at $T_{HC}\simeq10^{-6}$~GeV, will have current comoving strength~\cite{T1}
\begin{equation}
B_0\simeq 10^{-33} \left({M\over10^{17}}\right)^{2/3} \left({T_{RH}\over10^{10}}\right)^{1/3}\hspace{2mm}{\rm G}\,.  \label{B8}
\end{equation}
For typical values of the inflationary energy scale and the  reheat temperature (i.e.~when $M\simeq10^{17}$~GeV  and $T_{RH}\simeq10^{10}$~GeV), we obtain $B_0\simeq10^{-33}$~G. The latter, which is already twenty orders of magnitude larger than the typical conventional magnetic strengths quoted in the literature, can increase further to $\sim10^{-27}$~G by the time the galaxy is formed, especially when the more realistic scenario of an anisotropic protogalactic collapse is adopted~\cite{DBL}.

Although magnetic strengths of $B_0\sim10^{-27}$ are still outside the typical galactic-dynamo requirements, they are close enough to make one think that conventional models of inflationary magnetogenesis might still be able to work. For instance, turbulent motions can further increase the strength of the $B$-field, once the latter has been well inside the horizon. Another example, closer in spirit to our discussion here, is the possibility of a brief period of stiff-matter domination prior to the radiation era. During such a phase the $B$-field maintains constant magnitude~\cite{T1}, while the energy density of the dominant  matter component drops as $\rho_{SM}\propto a^{-6}$, leading to a sharp increase in the relative magnetic strength within a few expansion timescales. In particular, if $T_{SM}$ is the temperature at the end of the stiff-matter epoch, Eq.~(\ref{B8}) recasts into
\begin{equation}
B_0\simeq 10^{-33} \left({M\over10^{17}}\right)^{2/3} \left({T_{RH}\over10^{10}}\right)^{1/3}\left({T_{RH}\over T_{SM}}\right)^2\hspace{2mm}{\rm G}\,.  \label{B9}
\end{equation}
Consequently, assuming that stiff matter dominated the energy density of the universe between, say, $T_{RH}\simeq10^{10}$~GeV and $T_{SM}\simeq10^{7}$~GeV, would increase the magnetic strength from $10^{-27}$~G to $10^{-21}$~G by the time the galaxy is formed. The latter lies within the typical dynamo limits.

\section{Discussion}\label{sD}
All physical processes propagate at a finite speed, which ensures that local (causal) physics can never affect superhorizon perturbations. This is the root of the celebrated ``horizon problem''. Inflationary magnetic fields, like any other inflation-generated perturbation, can have coherence lengths vastly larger than the associated causal horizon. In addition, during inflation there are no electric currents and the aforementioned $B$-fields are not frozen into the matter, but they are ``free''. Since the post-inflationary magnetic-flux freezing is a causal process, mediated by the newly produced electric currents, it is always confined inside the horizon and can never affect $B$-fields with super-Hubble correlations. The opposite would had been a direct violation of the causality principle. Magnetic fields that are not frozen into the matter do not necessarily decay adiabatically. This means that superhorizon-sized magnetic fields do not need to obey the adiabatic ($B\propto a^{-2}$) decay-law, as long as the remain larger than the Hubble radius. Instead, the post-inflationary evolution of such fields is decided by the initial conditions at the start of the reheating phase. These are determined by the magnetic evolution during the de Sitter regime and by the specifics of the transition from inflation to reheating and then to the following epochs of radiation and dust. These transitions can be studied by appealing to Israel's junction conditions, which depend on whether the 3-dimensional hypersurface matching two successive cosmological epochs has zero or finite width, namely vanishing or thin surface layers.

We have outlined the basic features of three alternative and, to a large extent, complementary scenarios of initial conditions, referring the reader to~\cite{T1,T2} for further technical details. In all three cases, the initial conditions of the post-inflationary era have been calculated after assuming standard, purely exponential, de Sitter-type inflation. It is conceivable that adopting different inflationary models, such as power-law inflation for example, could add new features to the initial-condition paradigms discussed here and perhaps broaden their range.

The first of our scenarios assumes adiabatic magnetic decay during inflation and no thin shells on the transit hypersurface to reheating. Under these assumptions, the adiabatic decay of the $B$-field persists throughout its post-inflationary life. Therefore, for all practical purposes, this scenario reproduces the standard story of conventional primordial magnetogenesis, which leads to astrophysically irrelevant magnetic seeds. Scenario~B maintains the absence of surface layers on the matching hypersurfaces, but considers non-conventional magnetic fields that had been superadiabatically amplified during the de Sitter phase. We found that the superadiabatic magnetic amplification continues after inflation as well, at a rate determined by the specifics of the cosmological epoch. This is good news for the non-conventional mechanisms of primordial magnetogenesis, because a relatively mild enhancement of their $B$-fields during inflation will be capable of producing seeds that could feed the galactic dynamo today. In particular, the minimum required strength at the end of the de Sitter phase was found to lie around $10^{16}$~G. Such magnitudes are fairly straightforward to achieve through typical non-conventional mechanisms, without causing any backreaction or other known problems. There are good news for conventional magnetogenesis as well in scenario~C. Assuming that $B$-fields decayed adiabatically during the de Sitter expansion, but endowing the transition hypersurface to reheating with a thin surface layer, triggered the superadiabatic amplification of the aforementioned $B$-fields after inflation and for as long as they remain outside the Hubble radius. As a result, conventionally produced primordial magnetic fields can reach residual strengths much larger than those usually quoted in the literature, making one think that conventional magnetogenesis might still be able to work.\footnote{The three scenarios presented here have been approached from the relativistic point of view. As we have already mentioned, scenario~A is the standard model of conventional inflationary magnetogenesis, which has been discussed many times in the past and in a variety of ways. Scenarios~B and C, on the other hand, are new additions to the literature (see also~\cite{T1,T2}). Of these, scenario~B was recently reproduced in~\cite{C2}, both qualitatively and quantitatively, using an approach closer in spirit to high energy physics. Scenario~C could have been reproduced as well, if~\cite{C2} had also allowed for a ``jump'' in the value of the Hubble parameter (i.e.~set $[H_*]_-^+=H_*^+-H_*^-\neq0$) at the moment of the transition from inflation to reheating, in line with Israel's junction conditions.} Irrespective of whether this may prove to be the case or not, the underlying point is that by appealing to a basic physical principle, such as causality, and by ``exploiting'' the role of the initial conditions, one can introduce an alternative, fundamentally different and potentially pivotal approach to the question of cosmic magnetogenesis.

\end{document}